\documentclass{article}
\usepackage{amssymb,amsmath}

\if@twoside  m
\else
\fi \topmargin 5mm \headheight 0mm \headsep 0mm \textheight
225truemm \textwidth 150truemm
\newcommand{\D}{{\rm\ d}}                                             % Proof
\newtheorem{prop}{Proposition}                     % Proposition
\newtheorem{corol}{Corollary}                      % Corollary
\newtheorem{lemma}{Lemma}                          % Lemma
\newtheorem{rem}{Remark}                           % Remark

\newcommand{\qed}
           {\mbox{\quad\rule[-1.5pt]{.4em}{1.5ex}}}       % q.e.d.
\newcommand{\ie}{{\em i.e.}}
\newcommand{\eg}{{\em e.g.}}

\newcommand{\eps}{\varepsilon}

\begin{document}

\title{Curved planar quantum wires with Dirichlet and Neumann boundary
conditions}
\author{J.~Dittrich$^{a,b}$ and J.~K\v{r}\'\i\v{z}$^{a,c}$}
\date{}
\maketitle
\begin{quote}
{\small \em a) Nuclear Physics Institute, \\ \phantom{e)x}Academy
of Sciences of the Czech Republic, 250 68 \v Re\v z, Czech
Republic (mail address)\\
 b) Doppler Institute of Mathematical Physics,\\
\phantom{e)x}Faculty of Nuclear Sciences and Physical Engineering, Czech
 Technical University,\\
\phantom{e)x}B\v{r}ehov{\'a} 7, 115 19 Prague 1, Czech Republic \\
 c) Faculty of Mathematics and Physics, Charles University,\\
\phantom{e)x}V~Hole\v{s}ovi\v{c}k\'ach~2, 180 00 Prague 8, Czech
Republic
\\
 \rm \phantom{e)x}dittrich@ujf.cas.cz, kriz@ujf.cas.cz}
\vspace{8mm}

\noindent {\small We investigate the discrete spectrum of the
Hamiltonian describing a quantum particle living in the
two--dimensional curved strip.
We impose the Dirichlet and Neumann boundary condition on opposite
sides of the strip.
The existence of the discrete
eigenvalue below the essential spectrum threshold depends on the
sign of the total bending angle for the asymptotically straight
strips.}
\vspace {8mm}

\noindent{PACS numbers: 02.30.Tb, 02.30.Sa, 03.65.Db}
\end{quote}

%%%%%%%%%%%%%%%%%%%%%%%%%%%%%%%%%%%%%%%%%%%%%%%%%%%%%%%%%%%%%%%%%%%

\section{Introduction}

The spectral properties of curved quantum wires with Dirichlet boundary
condition were widely investigated (\eg\ \cite{ES}, \cite{GJ}, \cite{DE}).
It was shown that any small curvature of the tube in dimensions 2
and 3 produces at least one positive eigenvalue below the essential
spectrum threshold.
The problem of the existence of such eigenvalues in the straight quantum
waveguides with a combination of Dirichlet and Neumann boundary
conditions were also studied (\eg\ \cite{BGRS}, \cite{DK}).
In the present paper we consider curved planar quantum wires, where
Dirichlet boundary condition is imposed on one side of the wire, while
the Neumann boundary condition is imposed on the opposite side.
Both boundary conditions represent an impenetrable wall in the
sense that there is no current through the boundary.
They can model two types of interphases in a solid, \eg\ in a
superconductor, in principle.
It is worth to know whether the presence of two types of
interphases leads to new nanoscopic phenomena.

We prove that the existence of the discrete eigenvalue essentially
depends on the direction of the total bending of the strip.
Roughly speaking at least one bound state always exists if the Neumann
boundary condition is imposed on the "outer side" of the boundary, \ie\
the one which is locally longer.
On the other hand we show that there is no eigenvalue below the
essential spectrum threshold provided that the curvature does not change
its direction and the Neumann boundary condition is imposed on the inner
part of the boundary.

We consider a Schr\"odinger particle whose motion is confined to a
curved planar strip of the width $d$.
For definiteness, let a curve $\Gamma: \mathbb{R}\rightarrow\mathbb{R}^2$
be a $C^3$-diffeomorfism of the real axis onto $\Gamma(\mathbb{R})$.
Without loss of generality we can assume
$\dot\Gamma_1(s)^2+\dot\Gamma_2(s)^2=1$, so $s$ is the arc length of the
curve.
We define the normal vector $N$ to the curve $\Gamma$ and the signed
curvature $\gamma$ in the standard way $N_1(s)=-\dot\Gamma_2(s),\,
N_2(s)=\dot\Gamma_1(s),\, \gamma(s)=-\dot\Gamma(s)\cdot\dot N(s)$.
Let there exist real positive numbers $\gamma_-,\, \gamma_+$ such that
$-\gamma_-\leq\gamma(s)\leq \gamma_+<1/d$ for all $s\in
\mathbb{R}$ and let $\dot\gamma$ be bounded.
Let the strip which is the configuration space of the considered particle,
$\Omega\subset\mathbb{R}^2$, be defined by
$$\Omega=\bigl\{\langle x,y\rangle
\in \mathbb{R}^2\,\bigl |\,\langle x,y\rangle=\Gamma(s)+u N(s),\,s\in
\mathbb{R},\, u\in(0,d)\bigr \}.$$
We are going to consider only strips $\Omega$ which are not
selfintersecting.
We shall denote the parts of the boundary of the region $\Omega$ by
$\mathcal{D}=\bigl\{\langle x,y\rangle\in \mathbb{R}^2\, \bigl|\,
\langle x,y\rangle=\Gamma(s),\, s\in \mathbb{R}\bigr\}$ and $\mathcal{N}=\bigl\{\langle x,y\rangle\in \mathbb{R}^2\, \bigl|\,
\langle x,y\rangle=\Gamma(s)+d N(s),\, s\in \mathbb{R}\bigr \}$.
We impose the Dirichlet boundary condition on the set $\mathcal{D}$ and
the Neumann one on $\mathcal{N}$.

Putting $\hbar^2/2m=1$ we identify according to \cite{ES}, \cite{DE}, \cite{Davies}
the particle Hamiltonian with the unique self-adjoint operator acting in the
Hilbert space $L^2\bigl(\mathbb{R}\times(0,d)\, ,\,
\bigl(1-u\gamma(s)\bigr)\D s\D u\bigr)$ associated with the quadratic
form
\begin{equation}
\label{q0} q_0(f,g)=\int_\mathbb{R}\D s\int_0^d\D u\,\Biggl(
{1\over 1-u\gamma(s)}\overline{\partial f\over
\partial s}(s,u) {\partial g\over \partial
s}(s,u)+ \bigl (1-u\gamma(s)\bigr)
\overline{\partial f\over \partial u}(s,u){\partial g\over
\partial u}(s,u)\Biggr )
\end{equation}
defined on the domain
$$
Q(q_0)=\Bigl\{f \in H^1\bigl(\mathbb{R}\times(0,d)
\bigr)\, \Bigl | f(s,0)=0\,\, \rm{for\ a.\,e.\ } s \in \mathbb{R}\Bigr\},
$$
where
$H^1\bigl(\mathbb{R}\times(0,d)\bigr)$ is the
standard Sobolev space and $f(s,0)$ denotes the
trace of the function $f$ on the part of the boundary
$\mathbb{R}\times\{0\}$.
\begin{rem}
Using similar ``reflection" procedure like in the proof of Theorem
1 in \cite{DK} for trivial case of combination of Dirichlet and
Neumann boundary conditions we can see that our operator acts like
Laplace-Beltrami operator
\begin{equation}\label{Laplace-Beltrami}
L_\Omega^{DN}=-\Biggl({u\dot\gamma(s)\over
\bigl(1-u\gamma(s)\bigr)^3}{\partial\, \over \partial
s}+{1\over\bigl(1-u\gamma(s)\bigr)^2}{\partial^2\, \over
\partial s^2}+{\partial^2\, \over \partial u^2}-{\gamma(s)\over
1-u\gamma(s)} {\partial\, \over \partial u}\Biggr )
\end{equation}
on the domain $D(L_\Omega^{DN})=\Bigl\{f\in H^2\bigl(\mathbb{R}\times(0,d)
\bigr)\, \Bigl|\, f(s,0)=0, \, {\partial f\over \partial u}(s,d)=0
\,\, \rm{for\ a.\, e.\ } s\in \mathbb{R}\Bigr\}$.
\end{rem}
\section{Existence of the discrete eigenvalue}
\setcounter{equation}{0}
Our main goal is to prove three propositions on the existence,
resp. the absence, of the discrete spectrum in this, resp. the next
section.
We use the variational technique first introduced in \cite{GJ} for
proving the existence statements.
\begin{prop} \label{Ex1}
Let there exist a positive real number $s_0$ such that $\gamma(s)\leq
0$ for every $s$ satisfying $|s|\geq s_0$ and
$\int_{-s_0}^{s_0}\gamma(s)\D s <0$. Then $\inf
\sigma\bigl(L_\Omega^{DN}\bigr )<{\pi^2\over 4 d^2}$.
\end{prop}
\begin{PF}
For every $\Phi\in Q(q_0)$ we define the functional
\begin{equation}\label{q}
q[\Phi]=q_0(\Phi,\Phi)-{\pi^2\over4d^2}\|\Phi\|^2_{L^2\bigl(\mathbb{R}
\times (0,d),(1-u\gamma(s))\D u \D s\bigr)}.
\end{equation}
According to Rayleigh-Ritz variational method (see \eg\ \cite{Davies}) it
is enough to find a trial function $\Phi\in Q(q_0)$ such that
$q[\Phi]<0$ to prove our proposition.
We construct such trial function as follows.
Let $\varphi$ be an arbitrary function from the Schwartz space
$\mathcal{S}$ such that $\varphi(s)=1$ for $|s|\leq s_0$.
We use the external scaling, \ie\ we define a family of functions
\begin{equation}\label{phi_sigma}
\varphi_\sigma(s)\,=\,\left \lbrace\:\begin{array}{lll}
\varphi(s)&{\rm for}&|s|\leq s_0\\
\\
\varphi\bigl(\pm s_0 +\sigma (s\mp s_0)\bigr) & {\rm for} &
|s|\geq s_0
\end{array} \right .
\end{equation}
with upper signs for $s\geq s_0$, lower ones for $s\leq s_0$.
Now we take a function $\Phi_\sigma(s,u)=\sqrt{2\over d}\varphi_\sigma(s)
\sin{\pi u\over 2 d}$ as a trial function.
After straightforward calculation we obtain
\begin{eqnarray}\nonumber
q[\Phi_\sigma] &=& {2\over d}\int_{\mathbb{R}}\int_0^d\Biggl(
\Bigl|{d \varphi_\sigma(s) \over d s}\Bigr |^2{\sin^2{\pi u\over 2
d}\over 1-u\gamma(s)}\,+\,{\pi^2\over4d^2} \bigl(1-u \gamma(s)\bigr
)\, \bigl |\varphi_\sigma(s)\bigr |^2\cos {\pi u \over d} \Bigr )\D u \D
s \leq\\
\nonumber
&\leq&{\sigma \over 1-d\gamma_+}\,\|\dot\varphi
\|^2_{L^2(\mathbb{R})}\,+\, {1\over d}\int_{-s_0}^{s_0}\gamma(s) \D s.
\end{eqnarray}
The second term in this estimate is negative by assumption and it
is independent of $\sigma$.
Hence we can choose $\sigma$ so small that the whole functional is
negative, which finishes the proof. \qed
\end{PF}

If we assume in addition \eg\ that $\gamma$ has a compact
support, we can see using simple Dirichlet-Neumann bracketing
argument (see \eg\ \cite{RS}) that the essential spectrum begins at the value
$\pi^2/4d^2$.
So we can simply state the following
\begin{corol}
Let $\gamma$ have a compact support and $\int_\mathbb{R}\gamma(s)
\D s<0$.
Then the operator (\ref{Laplace-Beltrami}) has at least one
discrete eigenvalue.
\end{corol}
\begin{prop}\label{Ex2}
Let there exist a positive real number $s_0$ such that
$\gamma(s)=0$ for $|s|\geq s_0$.
Let $\int_{-s_0}^{s_0}\gamma(s)\D s=0$ and $\|\gamma\|_{L^2(
\mathbb{R})}>0$.
Then $\inf \sigma\bigl(L_\Omega^{DN}\bigr )<{\pi^2\over 4 d^2}$,
\ie\ there exists at least one positive discrete eigenvalue of
$L^{DN}$.
\end{prop}
\begin{PF}
We use the same technique as in the proof of the Proposition
\ref{Ex1}.
We only slightly modify the trial function.
Instead of (\ref{phi_sigma}) we define for any $\eps>0$ a family
of functions
\begin{equation}\label{phi_sigma,eps}
\varphi_{\sigma,\eps}(s)\,=\,\left \lbrace\:\begin{array}{lll}
\varphi(s)\bigl (1-\eps\gamma(s)\bigr)&{\rm for}&|s|\leq s_0\\
\\
\varphi\bigl(\pm s_0 +\sigma (s\mp s_0)\bigr) & {\rm for} & |s|\geq s_0
\end{array} \right .
\end{equation}
and $\Phi_{\sigma,\eps}(s,u)=\sqrt{2\over
d}\,\varphi_{\sigma,\eps}(s) \sin{\pi u\over 2 d}$.
We substitute this function into (\ref{q}) and we obtain
$$
q[\Phi_{\sigma,\eps}]\leq\sigma \|\dot\varphi
\|^2_{L^2(\mathbb{R})}\,+\,\eps^2{\|\dot\gamma
\|^2_{L^2(\mathbb{R})} \over 1-d\gamma_+}\,+\,\eps^2 {\|\gamma
\|^3_{L^3(\mathbb{R})} \over d}\,-\,\eps{2 \|\gamma
\|^2_{L^2(\mathbb{R})} \over d}.
$$
The term linear in $\eps$ is negative and choosing $\eps$
sufficiently small, we can make it dominating over the quadratic
one.
Finally, we fix this $\eps$ and choose a small enough $\sigma$ to
make the right hand side negative.\qed
\end{PF}
\begin{rem}
In fact we have used only one step of the original Goldstone and Jaffe
construction of the trial function in the proof of the Proposition
\ref{Ex1}, \ie\ the external scaling.
We usually need in addition to deform somehow this trial function inside
the interval, where original $\varphi(s)=1$.
In the present case this deformation is not necessary, because the first
transverse mode function itself is a good trial function due to the different
boundary conditions on both sides of the boundary.
This deformation is required for the case with zero bending angle,
which is seen in the proof of the Proposition \ref{Ex2}.

This situation is similar to the proof of Theorem 5.1 in
\cite{DEK}.
An unbounded curved quantum layer in $\mathbb{R}^3$ with
Dirichlet boundary condition is investigated there.
It is proved that there exists at least one isolated eigenvalue
provided the reference surface of the layer has a non-positive
total Gauss curvature.
If the total Gauss curvature is negative the proof does not
require to add a deformation term to the ``first transverse mode
function" while for the zero case it is necessary.
\end{rem}
\section{Absence of the discrete spectrum}
\setcounter{equation}{0}
The ``nonexistence" counterpart to Proposition \ref{Ex1} could be
formulated as follows.
Let there exist a positive real number $s_0$ such that
$\gamma(s)\geq 0$ for $s$ satisfying $|s|\geq s_0$ and
$\int_{-s_0}^{s_0}\gamma(s)\D s>0$,
then $\inf\sigma\bigl(L_\Omega^{DN}\bigr)\geq {\pi^2\over 4
d^2}$.
But this statement does not hold as the following example shows.
Let us consider the quantum wire with following properties.
There exist real numbers $s_1<s_2<s_3<s_4$ such that $\gamma(s)=0$
for $s\in (-\infty,s_1) \cup (s_2,s_3)  \cup (s_4,\infty)$,
$\gamma(s)<0$ on $(s_1,s_2)$ and $\gamma(s)>0$ on $(s_3,s_4)$.
According to our assumption $\int_{\mathbb{R}}\gamma(s)\D s>0$.
We take any function $\varphi(s) \in H^1( \mathbb{R})$ such that
$\varphi(s)=1$ on $(s_1,s_2)$ and $\varphi(s)=0$ on $(s_3,s_4)$.
Then we construct a function $\Phi=\sqrt{2\over
d}\,\varphi(s)\sin{\pi u\over 2 d}$.
We substitute this function into the functional (\ref{q}) and we
get
$$q[\Phi]=\|\dot\varphi\|^2_{L^2( \mathbb{R})}\,+\,{1\over
d}\,\int_{s_1}^{s_2}\gamma(s)\D s.$$
Hence $q[\Phi]<0$ for $d$ small enough, \ie\ for small width of
the strip, because $\gamma(s)<0$ on $(s_1,s_2)$ and the first term
does not depend on $d$.

We state the weaker proposition in the sequel, the nonexistence
statement is its direct corollary.
The proof of the proposition will use the following lemma.
\begin{lemma}\label{lem}
Let $\lambda_0(\alpha)$ be the lowest eigenvalue of the
self-adjoint operator $-{d^2 \over du^2}\,+\,V(u)$ acting in the
Hilbert space $L^2\bigl( [0,d] \bigr)$ with the domain
$D_\alpha=\bigl\{\psi \in
AC^2\bigl([0,d]\bigr)\,\bigl|\,\psi(0)=0,\,
\psi'(d)+\alpha\psi(d)=0\bigr\}$, where $\alpha$ is a real
number and $V$ is a real measurable bounded function on $[0,d]$.
Then for any $\alpha_1\geq\alpha_2$,\
$\lambda_0(\alpha_1)\geq \lambda_0(\alpha_2)$.
More precisely
$$
\lambda_0(\alpha_2)\leq
\lambda_0(\alpha_1)+(\alpha_2-\alpha_1){\psi_0(d)^2\over\|\psi_0\|^2},
$$
where $\psi_0$ is a real eigenfunction corresponding to
$\lambda_0(\alpha_1)$.
\end{lemma}
\begin{PF}
For $0<\varepsilon<d$ and $3+\alpha_2\varepsilon>0$
we define a family of functions
\begin{equation*}
\omega_\varepsilon(u)= \left \{
\begin{array}{lll}
0 & for & 0\leq u \leq d-\varepsilon\\
\\
{\alpha_1-\alpha_2\over
(3+\alpha_2\varepsilon)\varepsilon^2}(u-d+\varepsilon)^3\psi_0(d)
& for & d-\varepsilon\leq u \leq d.
\end{array}
\right .
\end{equation*}
Now let $\varphi_{0,\varepsilon}=\psi_0+\omega_\varepsilon$.
Then obviously $\varphi_{0,\varepsilon} \in C^2\bigl([0,d]\bigr)$,
$\varphi_{0,\varepsilon}(0)=0$ and
$\varphi'_{0,\varepsilon}(d)=-\alpha_2
\varphi_{0,\varepsilon}(d)$.
Hence $\varphi_{0,\varepsilon}\in D_{\alpha_2}$.
Using integration by parts and the definitions of
$\varphi_{0,\varepsilon}$ and $\omega_\varepsilon$ we directly
obtain
\begin{equation*}
\bigl(\varphi_{0,\varepsilon}\, , \,-\varphi''_{0,\varepsilon} + V
\varphi_{0,\varepsilon}\bigr)_{L^2\bigl([0,d]\bigr)}=
\bigl(\psi_0\, ,\, -\psi''_0 + V \psi_0\bigr)_{L^2\bigl([0,d]\bigr)} +
(\alpha_2 - \alpha_1) \psi_0(d)^2 + \mathcal{O}(\varepsilon)
\end{equation*}
and in the similar way
\begin{equation*}
\|\varphi_{0,\varepsilon}\|^2 = \|\psi_0\|^2 +
\mathcal{O}(\varepsilon^{3\over 2}).
\end{equation*}
According to Rayleigh-Ritz variational principle
\begin{equation*}
\lambda_0(\alpha_2)\leq {\bigl(\varphi_{0,\varepsilon}\, ,\, -\varphi''_{0,\varepsilon} + V
\varphi_{0,\varepsilon}\bigr)_{L^2\bigl([0,d]\bigr)} \over
\|\varphi_{0,\varepsilon}\|^2 } = {\lambda_0(\alpha_1)
\|\psi_0\|^2 + (\alpha_2-\alpha_1)\psi_0(d)^2 +
\mathcal{O}(\varepsilon) \over \|\psi_0\|^2 + \mathcal
{O}(\varepsilon^{3/2})}.
\end{equation*}
Finally, for $\varepsilon \rightarrow 0^+$ we have
\begin{equation*}
\lambda_0(\alpha_2)\leq
\lambda_0(\alpha_1)+(\alpha_2-\alpha_1){\psi_0(d)^2\over\|\psi_0\|^2}.
\qed
\end{equation*}
\end{PF}
\begin{rem}
The operator defined in the previous lemma is really self-adjoint,
it has the purely discrete spectrum and the eigenfunctions can be chosen
real \cite{BEH}.
\end{rem}
Now we are ready to prove the following proposition.
\begin{prop}\label{Neex}
Let $\gamma(s)\geq0$ for every $s \in \mathbb{R}$.
Then $\inf \sigma\bigl(L_\Omega^{DN}\bigr )\geq{\pi^2\over 4 d^2}$.
\end{prop}
\begin{PF}
We are going to show that for every $\Phi \in Q(q_0)$, the
functional $q[\Phi]$ is nonnegative.
We decompose the function $\Phi$ to the transverse orthonormal
basis, \ie\
$$
\Phi(s,u)\,=\,
\sum_{k=0}^\infty\phi_k(s)\chi_k(s;u),
$$
where $\chi_k(s;u)$ are normalized eigenfunctions of the self-adjoint
operator $h(s)$ acting in the Hilbert space
$L^2\bigl([0,d],(1-u\gamma(s))\D u\bigr)$
associated with the quadratic form
\begin{equation*}
\tilde q_0(f,g)=\int_0^d
\bigl (1-u\gamma(s)\bigr)
\overline{d f(s;u)\over d u}{d g(s;u)\over
d u} \D u
\end{equation*}
defined on the domain
$$
Q(\tilde q_0)=\bigl \{\psi \in AC\bigl([0,d]\bigr)\, \bigl |\,
\psi(s;0)=0 \bigr \}
$$
for almost every values of the parameter $s$.
Let us denote by $\lambda_k$ the eigenvalues of this operator
corresponding to eigenfunctions $\chi_k$.
Then we can write the functional $q$ in the following form
\begin{eqnarray}\nonumber
q[\Phi]&=&\int_\mathbb{R}\D
s\int_0^d\D u\Bigl (\Bigl |{\partial \Phi(s,u) \over \partial s} \Bigr|^2 {1\over
1-u\gamma(s)}
\,+\, \sum_{k=0}^\infty|\phi_k(s)\chi_k(s;u)|^2 \Bigl (\lambda_k-{\pi^2\over 4 d^2}\Bigr )\bigl
(1-u\gamma(s)\bigr)\Bigr)\geq\\
\nonumber
&\geq&\sum_{k=0}^\infty\int_\mathbb{R}\D s\int_0^d\D u\,
|\phi_k(s)\chi_k(s;u)|^2 \Bigl (\lambda_k-{\pi^2\over 4 d^2}\Bigr )\bigl
(1-u\gamma(s)\bigr).
\end{eqnarray}
So it is enough to show that the lowest eigenvalue $\lambda_0(s)$ of the operator
$h(s)$ satisfies the condition $\lambda_0\geq {\pi^2\over 4 d^2}$.
It is easy to see that
$
h(s)=-{d^2\over d u^2} + {\gamma(s)\over 1-u\gamma(s)}\,{d\over d u}
$
with the domain \cite{BEH} $$D\bigl(h(s)\bigr)=\bigl\{\psi \in AC^2\bigl ([0,d]\bigr)\,
\bigl |\, \psi(s;0)=0,\, \psi'(s;d)=0\bigr \}.$$
Therefore $\chi_k(s;\cdot)$ are solutions of the equation
\begin{equation}\label{lambda}
-\chi''(s;u) + {\gamma(s)\over 1-u\gamma(s)}\, \chi'(s;u) =
\lambda(s) \chi(s;u).
\end{equation}
They are linear combinations of Bessel functions
$$
\chi(s;u) = A(s)\, J_0\biggl ({\sqrt{\lambda(s)}\over
\gamma(s)}\, \bigl( 1-u \gamma(s)\bigr )\biggr )\, +\, B(s)\, N_0\biggl ({\sqrt{\lambda(s)}\over
\gamma(s)}\, \bigl( 1-u \gamma(s)\bigr )\biggr ).
$$
Imposing the prescribed boundary conditions we obtain the equation
for $\lambda(s)$.
So $\lambda_0(s)$ is the lowest solution of the equation
\begin{equation}\label{lambda}
J_0\biggl({\sqrt{\lambda(s)}\over \gamma(s)}\biggr)\,N_1\biggl({\sqrt{\lambda(s)}\over
\gamma(s)}\, \bigl( 1-d \gamma(s)\bigr)\biggr)-
N_0\biggl ({\sqrt{\lambda(s)}\over \gamma(s)}\biggr)\,
J_1\biggl ({\sqrt{\lambda(s)}\over
\gamma(s)}\, \bigl( 1-d \gamma(s)\bigr )\biggr )= 0,
\end{equation}
which will be used later.

Unitary transformation $\psi(s;u) = \sqrt{1-u\gamma(s)}\,
\chi(s;u)$ from $L^2\Bigl([0,d],\bigl(1-u\gamma(s)\bigr)\D
u\Bigr)$ onto $L^2\bigl([0,d],\D u\bigr)$
transforms our problem to the one of searching for
the lowest eigenvalue of the operator
$$
\tilde h(s) = -{ d^2\over d u^2} - {\gamma(s)^2\over 4\bigl (1-u
\gamma(s)\bigr)^2}
$$
on the domain
\begin{equation}\label{Dh}
D\bigl(\tilde h(s)\bigr)=\bigl\{\psi \in AC^2\bigl ([0,d]\bigr)\,
\bigl |\, \psi(s;0)=0,\, \psi'(s;d)+{\gamma(s)\over 2\bigl(1-d\gamma(s)\bigr)}
\psi(s;d)=0\bigr\}.
\end{equation}
Let us introduce two more operators acting in the Hilbert space
$L^2\bigl([0,d]\bigr)$,
$h_0=-{d^2\over du^2}$ with
the domain $D\bigl(h(s)\bigr)$ and $h_1(s)=-{d^2\over du^2}$ with
the domain $D\bigl(\tilde h(s)\bigr)$.
It is known that the lowest eigenvalue of $h_0$ is ${\pi^2\over 4
d^2}$.
Let the lowest eigenvalue of $h_1(s)$ is denoted by
$\lambda_{0,1}(s)$ and the corresponding real eigenfunction by $\psi_0$.
The normalized eigenfunction corresponding to the lowest
eigenvalue $\lambda_0(s)$ of the operator $\tilde h(s)$ is denoted by
$\psi_1$.
Using Lemma \ref{lem} and the Rayleigh-Ritz variational principle
we can write
\begin{eqnarray}\nonumber
{\pi^2\over 4 d^2} &\leq& \lambda_{0,1}(s) - {\gamma(s)\over 2\bigl(
1-\gamma(s)\bigr)}\, {\psi_0(s;d)^2\over \|\psi_0\|^2}\leq\\
\nonumber\\
\label{odhadlambda}
&\leq&\lambda_0(s)- {\gamma(s)\over 2\bigl(
1-\gamma(s)\bigr)}\, {\psi_0(s;d)^2\over \|\psi_0\|^2} + \int_0^d
{\gamma(s)^2\over 4\bigl (1-u\gamma(s)\bigr)^2}
\bigl|\psi_1(s;u)\bigr |^2 \D u \leq\\
\nonumber\\
\nonumber
&\leq& \lambda_0(s)- {\gamma(s)\over 2\bigl(
1-\gamma(s)\bigr)}\, {\psi_0(s;d)^2\over \|\psi_0\|^2} +
{\gamma(s)^2\over 4\bigl (1-d\gamma(s)\bigr)^2}.
\end{eqnarray}
The eigenfunction $\psi_0$ is given by the equation
\begin{equation*}
-\psi''_0(s;u)=\lambda_{0,1}(s)\,\psi_0(s;u)
\end{equation*}
and it satisfies boundary conditions defined in (\ref{Dh}).
Hence $\psi_0(s;u)=\sin\sqrt{\lambda_{0,1}(s)}u$ and $\lambda_{0,1}(s)$ is
the lowest solution of the equation
$$
\sqrt{\lambda_{0,1}(s)}\,=\,-\,\alpha(s)\,\tan \sqrt{\lambda_{0,1}(s)}d
$$
where $\alpha(s)={\gamma(s)\over 2\bigl (1-d\gamma(s)\bigr)}$.
Now necessarily $\lambda_{0,1}(s)\in \Bigl({\pi^2\over 4 d^2},{\pi^2\over
d^2}\Bigr)$ in accordance with the first estimate in
(\ref{odhadlambda}).
We can calculate
\begin{equation*}
{\psi_0(s;d)^2\over \|\psi_0\|^2}={\sin^2\sqrt{\lambda_{0,1}(s)}d
\over {d\over 2}-{\sin 2\sqrt{\lambda_{0,1}(s)}d\over 4
\sqrt{\lambda_{0,1}(s)}}}\, =\, {2\lambda_{0,1}(s)\over
d\bigl(\alpha(s)^2+\lambda_{0,1}(s)\bigr) + \alpha(s)}.
\end{equation*}
To see that $\lambda_0(s)\geq {\pi^2\over 4 d^2}$, which we want
to prove, it is now enough to show
\begin{equation*}
\alpha(s)\,{2\lambda_{0,1}(s)\over d\bigl(\alpha(s)^2+\lambda_{0,1}(s)\bigr) +
\alpha(s)}\,-\,\alpha(s)^2 \geq 0
\end{equation*}
due to (\ref{odhadlambda}).
Using inequality $\lambda_{0,1}(s)\geq {\pi^2\over 4 d^2}$ it can
be seen that the inequality above holds provided
\begin{equation}\label{odhad}
{\pi^2\over 4} \geq {\gamma(s)^2d^2 \over
4\bigl(1-d\gamma(s)\bigr)^2}\, {1+{d\gamma(s)\over
2\bigl(1-d\gamma(s)\bigr)} \over 2-{d\gamma(s)\over
2\bigl(1-d\gamma(s)\bigr)}}
\end{equation}
and $d\gamma(s)<{4\over5}$.
The right hand side of the inequality (\ref{odhad}) is obviously increasing
function of $d\gamma(s)$ and it reaches value $0$ for
$d\gamma(s)=0$ and value $2<{\pi^2\over 4 }$ for
$d\gamma(s)={2\over3}$.

In the remaining interval $d\gamma(s)\in [{2\over 3},1)$ we can study directly the
equation (\ref{lambda}) using up to the first four terms of the known expansions of
Bessel functions into series \cite{W}.
It can be seen that the equation cannot be satisfied for values
$d\gamma(s)\in [{2\over 3},1)$ and $\lambda_0< {\pi^2\over 4
d^2}$.
\qed
\end{PF}\linebreak
\begin{corol}
Let $\gamma$ satisfy the same assumption as in Proposition
\ref{Neex} and let it have a compact support.
Then the operator (\ref{Laplace-Beltrami}) has no discrete
eigenvalue.
\end{corol}
\section*{Acknowledgements}
The authors thank Professor Pavel Exner and Professor Pierre
Duclos for  useful discussions.
The work is supported by GA ASCR grant IAA 1048101.


\begin{thebibliography}{article}
   \bibitem {ES}
   P.~Exner, P.~\v Seba: ``Bound states in curved waveguides", J.
   Math. Phys. {\bf 30}, 2574--2580 (1989).
   \vspace{-.8em}
   \bibitem{GJ}
   J.~Goldstone, R.L.~Jaffe: ``Bound states in twisting tubes",
   Phys. Rev. {\bf B45}, 14100--14107 (1992).
   \vspace{-.8em}
   \bibitem{DE}
   P.~Duclos, P.~Exner: ``Curvature-induced bound states in quantum
   waveguides in two and three dimensions", Rev.Math.Phys. {\bf
   7}, 73--102(1995).
   \vspace{-.8em}
   \bibitem{BGRS}
   W.~Bulla, F.~Gesztesy, W.~Renger, B.~Simon: ``Weakly coupled
   bound states in quantum waveguides", Proc. Amer. Math. Soc.
   {\bf 127}, 1487--1495 (1997).
   \vspace{-.8em}
   \bibitem{DK}
   J.~Dittrich, J.~K\v r\'\i \v z: ``Bound states in straight quantum
   waveguides with combined boundary condition", e-preprint
   mp\_arc~01--458, submitted to J. Math. Phys.
   \vspace{-.8em}
   \bibitem{Davies}
   E.B.~Davies: {\em Spectral Theory and Differential Operators},
   University Press, Cambridge 1995.
   \vspace{-.8em}
   \bibitem{RS}
   M.~Reed, B.~Simon: {\em  Methods of Modern Mathematical Physics,
IV.Analysis of Operators}, Academic Press, New York 1978.
   \vspace{-.8em}
   \bibitem{DEK}
   P.~Duclos, P.~Exner, D.~Krej\v ci\v r\'\i k:``Bound States in
   Curved Quantum Layers", Commun. Math. Phys. {\bf 223}, 13--28
   (2001).
   \vspace{-.8em}
   \bibitem{BEH}
   J.~Blank, P.~Exner, M.~Havl\'\i \v cek: {\em Hilbert Space Operators
   in Quantum Physics}, AIP, New York 1994.
   \vspace{-.8em}
   \bibitem{W}
   G.N.~Watson: {\em A Treatise on the Theory of Bessel
   functions}, University Press, Cambridge 1952.
\end{thebibliography}
\end{document}